\documentstyle[epsf]{mn}

\newif\ifAMStwofonts
\def\etal{{et al. \rm}}
\def\FE{[Fe II] $\lambda$1.644 $\mu$m}
\def\FES{[Fe II] $\lambda$1.644 $\mu$m }
\def\fe{[Fe II] }
\def\ls{$L$(\FE) }

\def\lf{$L_{{\rm [Fe \: II]}}$}
\def\lfs{$L_{{\rm [Fe \: II]}}$ }
\def\s{SNRs}
\ifoldfss
  \ifCUPmtlplainloaded \else
    \NewTextAlphabet{textbfit} {cmbxti10} {}
    \NewTextAlphabet{textbfss} {cmssbx10} {}
    \NewMathAlphabet{mathbfit} {cmbxti10} {} % for math mode
    \NewMathAlphabet{mathbfss} {cmssbx10} {} %  "   "    "
  \fi
  \ifAMStwofonts
    \ifCUPmtlplainloaded \else
      \NewSymbolFont{upmath} {eurm10}
      \NewSymbolFont{AMSa} {msam10}
      \NewMathSymbol{\upi}     {0}{upmath}{19}
      \NewMathSymbol{\umu}     {0}{upmath}{16}
      \NewMathSymbol{\upartial}{0}{upmath}{40}
      \NewMathSymbol{\leqslant}{3}{AMSa}{36}
      \NewMathSymbol{\geqslant}{3}{AMSa}{3E}

    \fi
  \fi
\fi % End of OFSS

\ifnfssone
  \newmathalphabet{\mathit}
  \addtoversion{normal}{\mathit}{cmr}{m}{it}
  \addtoversion{bold}{\mathit}{cmr}{bx}{it}
  \newmathalphabet{\mathbfit} % math mode version of \textbfit{..}
  \addtoversion{normal}{\mathbfit}{cmr}{bx}{it}
  \addtoversion{bold}{\mathbfit}{cmr}{bx}{it}
  \newmathalphabet{\mathbfss} % math mode version of \textbfss{..}
  \addtoversion{normal}{\mathbfss}{cmss}{bx}{n}
  \addtoversion{bold}{\mathbfss}{cmss}{bx}{n}
  \ifAMStwofonts
    \ifCUPmtlplainloaded \else
      \UseAMStwoboldmath
      \makeatletter
      \new@mathgroup\upmath@group
      \define@mathgroup\mv@normal\upmath@group{eur}{m}{n}
      \define@mathgroup\mv@bold\upmath@group{eur}{b}{n}
      \edef\UPM{\hexnumber\upmath@group}
      \new@mathgroup\amsa@group
      \define@mathgroup\mv@normal\amsa@group{msa}{m}{n}
      \define@mathgroup\mv@bold\amsa@group{msa}{m}{n}
      \edef\AMSa{\hexnumber\amsa@group}
      \makeatother
      \mathchardef\upi="0\UPM19
      \mathchardef\umu="0\UPM16
      \mathchardef\upartial="0\UPM40
      \mathchardef\leqslant="3\AMSa36
      \mathchardef\geqslant="3\AMSa3E
    \fi
  \fi
\fi % End of NFSS release 1

\ifnfsstwo
  \DeclareMathAlphabet{\mathbfit}{OT1}{cmr}{bx}{it}
  \SetMathAlphabet\mathbfit{bold}{OT1}{cmr}{bx}{it}
  \DeclareMathAlphabet{\mathbfss}{OT1}{cmss}{bx}{n}
  \SetMathAlphabet\mathbfss{bold}{OT1}{cmss}{bx}{n}
  \ifAMStwofonts
    \ifCUPmtlplainloaded \else
      \DeclareSymbolFont{UPM}{U}{eur}{m}{n}
      \SetSymbolFont{UPM}{bold}{U}{eur}{b}{n}
      \DeclareSymbolFont{AMSa}{U}{msa}{m}{n}
      \DeclareMathSymbol{\upi}{0}{UPM}{"19}
      \DeclareMathSymbol{\umu}{0}{UPM}{"16}
      \DeclareMathSymbol{\upartial}{0}{UPM}{"40}
      \DeclareMathSymbol{\leqslant}{3}{AMSa}{"36}
      \DeclareMathSymbol{\geqslant}{3}{AMSa}{"3E}
    \fi
  \fi
\fi % End of NFSS release 2

\ifCUPmtlplainloaded \else
  \ifAMStwofonts \else % If no AMS fonts
    \def\upi{\pi}
    \def\umu{\mu}
    \def\upartial{\partial}
  \fi
\fi

\title[Near-IR \fe emission from SNRs]{Near-infrared \fe emission from supernova remnants and the supernova rate of starburst galaxies}
\author[T. Morel, R. Doyon and N. St-Louis]
{\parbox{179mm}{\begin{flushleft}
\vspace{-0.5cm}
{\LARGE T. Morel,$^{1, 2, 3}$} 
\thanks{e-mail: morel@iucaa.ernet.in}
{\LARGE R. Doyon$^3$}
{\LARGE and N. St-Louis$^3$}
\end{flushleft}
}\vspace*{0.200cm}\\  
\parbox{159mm}{
$^1$ Inter-University Centre for Astronomy and Astrophysics (IUCAA), Post Bag 4, Ganeshkhind, Pune, 411 007, India\\
$^2$ Astrophysics Group, Imperial College of Science, Technology and Medicine, Blackett Laboratory, Prince Consort Road, London SW7 2BZ\\
$^3$ D\'epartement de Physique, Universit\'e de Montr\'eal, 
C. P. 6128, Succ. Centre-Ville, Montr\'eal, Qu\'ebec, Canada, H3C 3J7; and
Observatoire du Mont M\'egantic}}

\date{Accepted ???.
      Received ???;
      in original form ??? }

\pagerange{\pageref{firstpage}--\pageref{lastpage}}
\pubyear{2001}

\begin{document}

\maketitle

\label{firstpage}

\begin{abstract}

In an effort to better calibrate the supernova rate of starburst galaxies as determined from near-infrared \fe features, we report on a \FES line-imaging survey of a sample of 42 optically-selected supernova remnants (SNRs) in M33. A wide range of \fe luminosities are observed within our sample (from less than 6 to 695 L$_{\odot}$). Our data suggest that the bright \fe SNRs are entering the radiative phase and that the density of the local interstellar medium (ISM) largely controls the amount of \fe emission. We derive the following relation between the \FES line luminosity of {\em radiative} SNRs and the electronic density of the postshock gas, $n_e$: \lfs (L$_{\odot}$) $\approx$ 1.1 $n_e$ (cm$^{-3}$). We also find a correlation in our data between \lfs and the metallicity of the shock-heated gas, but the physical interpretation of this result remains inconclusive, as our data also show a correlation between the metallicity and $n_e$. The dramatically higher level of \fe emission from SNRs in the central regions of starburst galaxies is most likely due to their dense environments, although metallicity effects might also be important. The typical [Fe II]-emitting lifetime of a SNR in the central regions of starburst galaxies is found to be of the order of 10$^4$ yr. On the basis of these results, we provide a new empirical relation allowing the determination of the current supernova rate of starburst galaxies from their integrated near-infrared \fe luminosity. 
\end{abstract}


\section{Introduction}

Early near-infrared (IR) spectroscopic observations of galactic SNRs have shown that these objects present remarkably
 strong \fe features (e.g. Graham, Wright \& Longmore 1987). This has been proposed to be mainly due to the existence of an extended postshock region in which the ionization conditions are such that Fe$^+$ can be efficiently excited by electron collisions (Mouri, Kawara \& Taniguchi 2000). In contrast, the spatial extent of this zone, where hydrogen is partially ionized, 
is small in photoionized regions, resulting in little emission (e.g. Luhman, Engelbracht
\& Luhman 1998).

This peculiarity can be used as a diagnostic probe of supernova activity in distant starburst galaxies where the SNRs are generally unresolved and are thought to produce much of the integrated near-IR \fe emission (e.g. Vanzi \& Rieke 1997). Of particular interest is the fact that using near-IR \fe lines to estimate supernova activity can potentially be superior to the usual radio technique, in that it more clearly distinguishes H II regions from SNRs. 

From
estimates of the {\em total} \fe luminosity of a starburst
galaxy (${\cal L}_{{\rm [Fe \: II]}}$) and assumptions regarding the typical values of the \fe luminosity ($L_{{\rm [Fe \: II]}}$) and [Fe II]-emitting
lifetime ($t_{{\rm [Fe \: II]}}$) of a {\em single} SNR, one can define the supernova
rate, $\eta$, as:
\begin{equation}
\eta  = \frac{{\cal L}_{{\rm [Fe \: II]}}}{t_{{\rm [Fe \: II}]} \:  L_{{\rm [Fe \: II]}}} {\rm \hspace*{0.2cm}  yr^{-1}}
\end{equation}
This quantity can be used, for instance, to set constraints on the star formation history of starburst galaxies (e.g. Leitherer \& Heckman 1995; Kotilainen \etal 1996). 

Unfortunately, the supernova rate derived by this technique is considerably uncertain, as it scales
with $t_{{\rm [Fe \: II]}}$ and $L_{{\rm [Fe \: II]}}$ which are poorly
known. The [Fe II]-emitting lifetime
of a SNR is generally taken to be of the order of 1--2 $\times$ 10$^4$ yr, but this estimate lies on weak grounds (e.g. van der Werf \etal 1993). On the other hand, the few \FES luminosities quoted in the
literature for SNRs in our Galaxy, the LMC or M33 span a wide range of values (from 0.3 to 720
L$_\odot$; Graham \etal 1987; Oliva, Moorwood \&
Danziger 1989, 1990; Lumsden \& Puxley 1995 -- hereafter LP). One issue of particular importance is to understand
 why SNRs in starburst galaxies exhibit dramatically higher \fe luminosities (up to  1.6 $\times$ 10$^5$ L$_\odot$ in M82; Greenhouse \etal 1997). One order of magnitude higher values are also observed in NGC 253 (Forbes \etal 1993). Before using near-IR \fe features to estimate supernova activity in star-forming galaxies that presumably exhibit widely different properties (e.g. metallicity), one must first understand what parameters control the \fe properties of individual SNRs. Many factors (e.g. evolutionary status of the SNR, density or metallicity of the local ISM) are susceptible to play a
role in producing this observed luminosity range,
although their relative importance has yet to be established. 

\section{A near-IR \fe line-imaging survey of extragalactic \s}
In an effort to address this issue, we present the results of a project in which narrowband images are used to secure \FES luminosities for a large sample
of extragalactic SNRs spanning a wide range of physical properties and evolutionary status. Previous efforts to determine the \fe luminosities of individual SNRs have mainly concentrated on aperture photometry of galactic remnants. The shortcomings of such an approach are numerous: (i) those observations sample, in most cases, only a very small part of the SNR, rendering the total value of the \fe luminosity considerably uncertain; (ii) because of the different apertures used, the published \fe line luminosities do not
constitute a homogeneous database based on which meaningful statistical
studies can be conducted; (iii) the \fe luminosities are entached of large uncertainties, as the distance to these SNRs and the extinction corrections are generally poorly known. On the other hand, because of the unavoidable limited size of the
aperture, surveys in nearby galaxies carried out using long-slit spectroscopy exclude the largest objects and are therefore
biased towards the youngest SNRs (see LP). Line-imaging observations of SNR populations in local group galaxies do not suffer from these limitations.

M33 is an obvious target in this respect because of its relatively face-on orientation ($i$ $\approx$ 55$^{\circ}$; Garcia-Gomez \& Athanassoula 1991), well-known distance ($D$ $\approx$ 840 kpc; Freedman, Wilson \& Madore 1991) and substantial abundance gradient that allows an investigation of metallicity effects (Smith \etal 1993). Furthermore, it has the largest catalogued extragalactic SNR population (Gordon \etal 1999). Of particular importance is the fact that a large body of observational data exists for SNRs in this galaxy; as will be shown below, this will allow us to relate the near-IR \fe emission to 
observations in other parts of the electromagnetic spectrum. The latest catalogue of optically-selected SNR candidates in M33 comprises 98 objects (Gordon \etal 1998); many of which being detected at radio frequencies (Gordon \etal 1999, and references therein). Among this sample, 72 objects have been spectroscopically confirmed and therefore also have well-defined optical-line properties (Gordon \etal 1998, and references therein). In addition, a {\em ROSAT} survey of M33 has associated some X-ray sources with known SNRs (Long et al. 1996). 

\section[]{Observations and Reduction Procedure}
Our observations were obtained at the Canada-France-Hawaii Telescope (CFHT) in 1997 and 1998, using the IR
cameras MONICA and REDEYE, respectively. For both runs, the cameras were equipped
with a 256 $\times$ 256 pixels HgCdTe NICMOS-3 detector array. At the {\it f}/8
Cassegrain focus of the CFHT, the image scales are 0.246 (MONICA) and 0.5 arcsec pixel$^{-1}$ (REDEYE). For the assumed distance to M33 ($D$ $\approx$ 840 kpc),
this translates into a physical scale of 1 and 2 pc pixel$^{-1}$, respectively. In
this configuration, the fields of view are about 63 (MONICA) and 128 arcsec (REDEYE). The interference filters used are a narrowband filter (FWHM $\approx$ 0.02 $\mu$m) centred on \FES (a$^4$D$_{7/2}$ $\rightarrow$ a$^4$F$_{9/2}$) and a wideband {\em
H} filter. The \fe filter is sufficiently wide that the broadening due to
motions of the remnant gas, the rotation or the redshift
of M33 do not shift lines to significantly different parts of the
transmission curve. In total 42 objects drawn from the optically-selected catalogue of Gordon \etal (1998) have been observed. The vast majority has been spectroscopically confirmed (Blair \& Kirshner 1985; Smith \etal 1993; Gordon \etal 1998). Particular attention has been paid to observe SNRs with widely different properties. The journal of observations is presented in Table 1. 

For each SNR we have obtained 5 images: one centred on the object and 4 offset to the NO, NE, SO and SE by 22 and 45 arcsec for MONICA and REDEYE, respectively. These images were later combined in a mosaic centred on the SNR. Two type of domeflats have been obtained (dome lights ``on'' and ``off''). The latter has been subtracted from the former in order to remove the contribution from the thermal background. These domeflats have been used to correct for changes in the quantum efficiency of the pixels in each object image. A mean sky frame has been created by applying a pixel-to-pixel median filtering procedure with proper masking of sources to the data frames themselves. No off-source images have been obtained. This procedure is justified in our case, as we are not attempting to detect uniform large-scale emission, but rather compact sources on the arcsecond scale (see Fig.1). After subtraction of this mean sky frame to the individual object frames and multiplication by a bad pixel mask in order to account for dead pixels, the object images were aligned and coadded. The astrometric calibration was carried out with reference to positions of stars in the USNO-A2.0 or GSC1.2 catalogues. The positions are accurate to within 3 arcsec. 

\clearpage

\begin{figure}
\epsfxsize=24cm
\epsfysize=24cm
\epsfbox{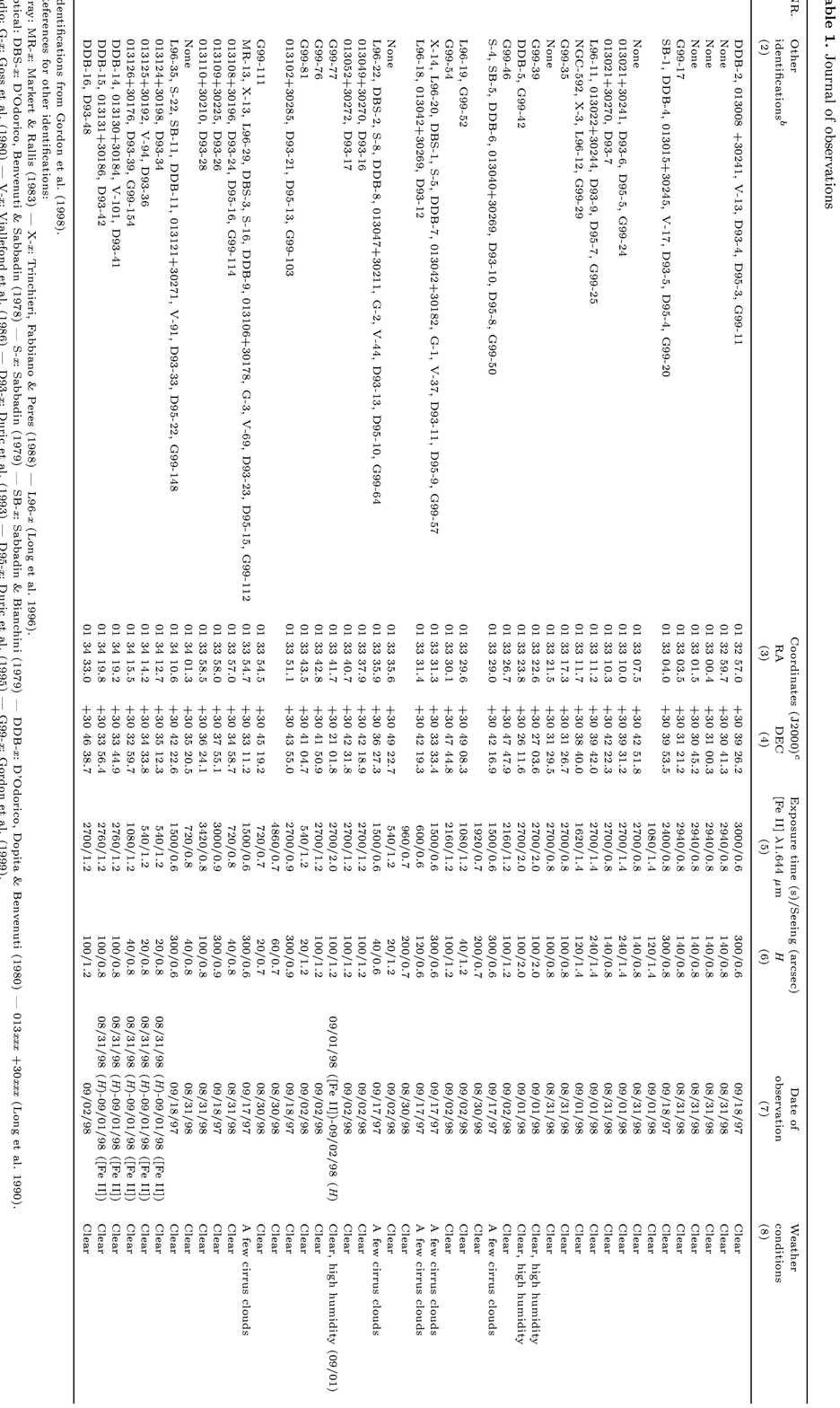}
\end{figure}

\clearpage

\begin{figure}
\epsfxsize=22cm
\epsfysize=28cm
\epsfbox{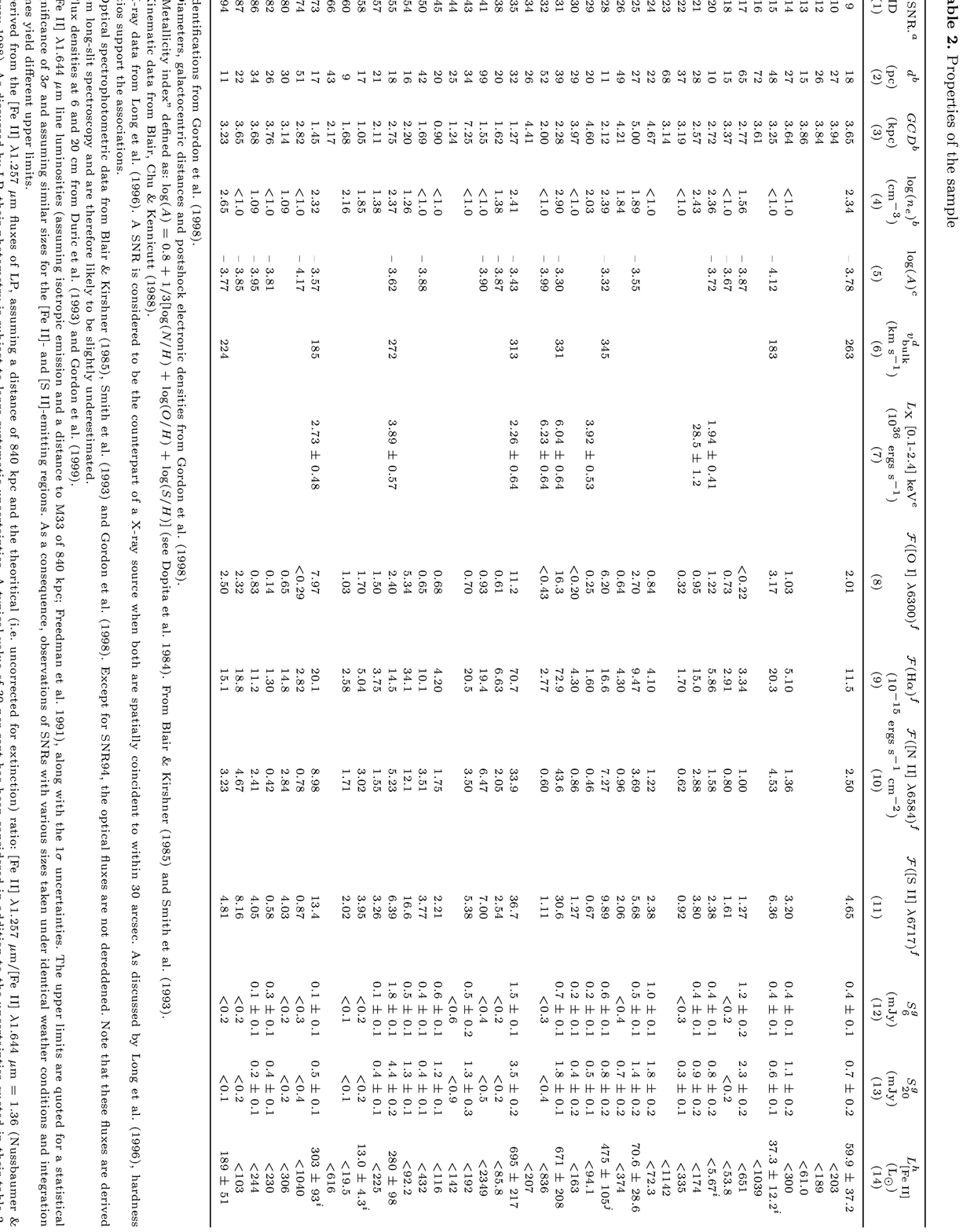}
\end{figure}

\clearpage

\section{Overview of the survey}
Among our sample of 42 objects, only 7 have been firmly detected (we address in Section 6.1 the physical reasons for such a low detection rate). Fig. 1 shows the \fe and {\em H}-band images of this subset. The \FES luminosities are quoted in Table 2, along with the 3$\sigma$ upper limits for the remaining SNRs (col. [14]). The targets were observed at similar airmasses as the standard stars, eliminating the need to correct for atmospheric extinction. This is legitimate in the {\em H} band, where such corrections are small. Using the optical extinction measurements of Blair \& Kirshner (1985) and the interstellar reddening law of Rieke \& Lebofsky (1985), it is also found that reddening corrections are negligible. The contribution of Br12 $\lambda$1.641 $\mu$m and [Si I] $\lambda$1.645 $\mu$m to the total flux in the \fe filter has been neglected (see Oliva \etal 1989). The non-detection of any significant flux in the {\em H}-band images also indicates that the contribution from continuum emission is negligible (see also LP). The flux calibration was achieved by using faint standard stars in the list of Hunt \etal (1998). The internal accuracy is typically 0.011 mag in the {\em H} band.

\begin{figure*}
\epsfxsize=14cm
\epsfysize=14cm
\epsfbox{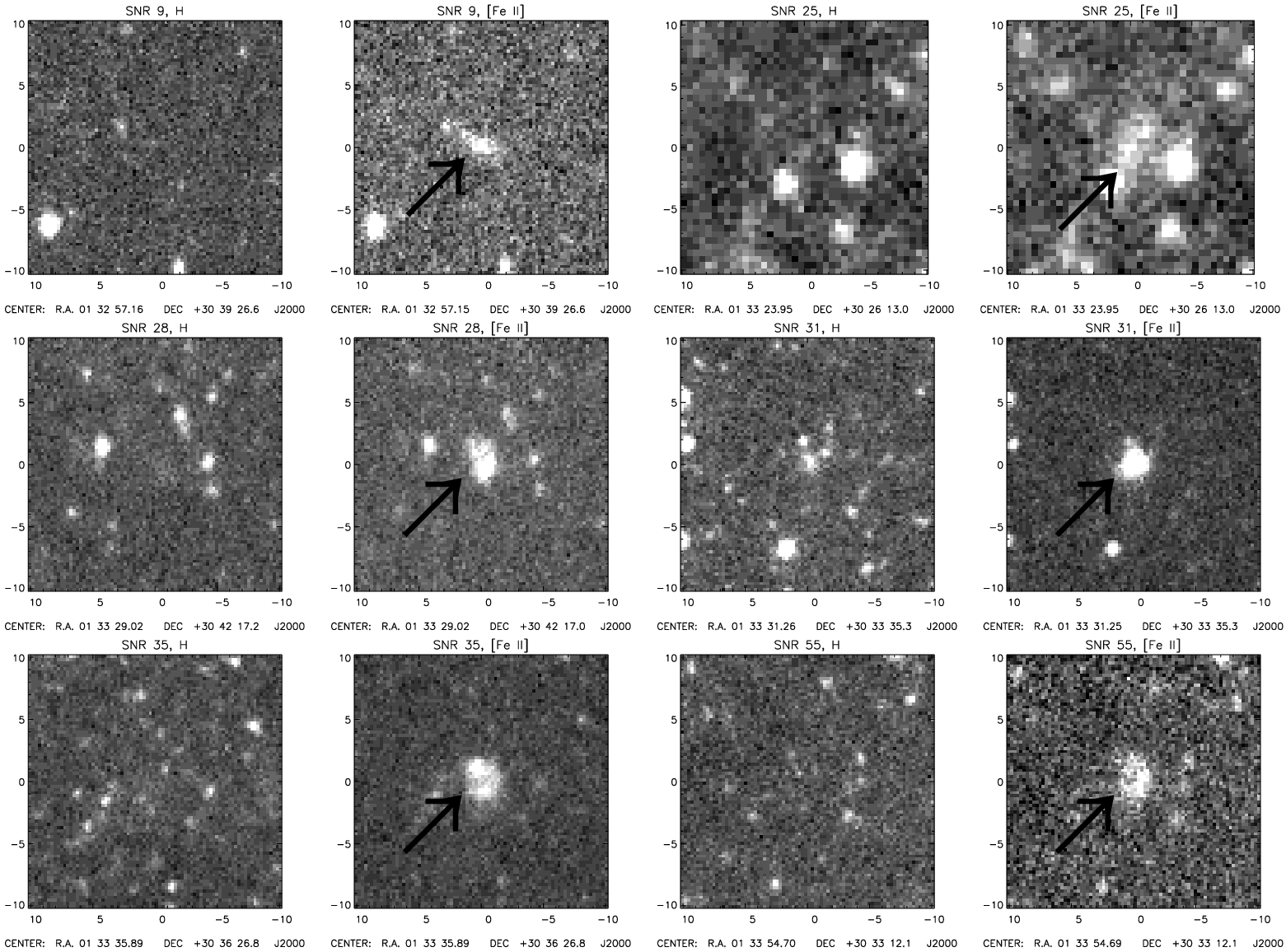}
\end{figure*}

\begin{figure*}
\epsfxsize=14cm
\epsfysize=14cm
\epsfbox{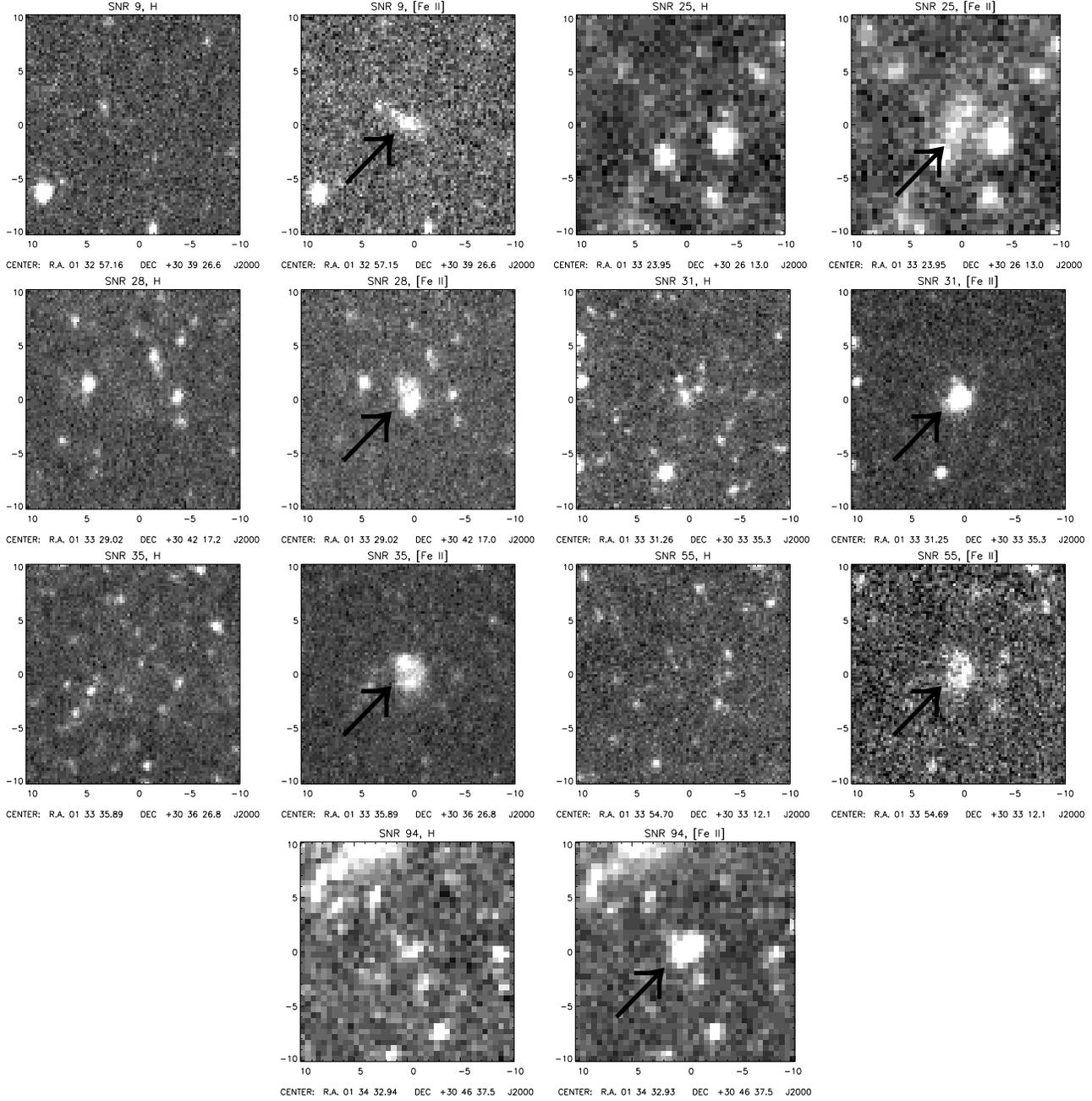}
\vspace*{-10.7cm}
\caption{Comparison of the $H$-band ({\em left panels}) \rm and \FES ({\em right panels}) \rm images of the detected SNRs. The identification of the SNR (taken from Gordon \etal 1998) is indicated on top of each panel. The location of the SNRs is indicated by an arrow. At the assumed distance to M33, 1 arcsec corresponds to 4.1 pc. The field of view is 20 arcsec $\times$ 20 arcsec. North is up and east is to the left.}
\end{figure*}

Some of our objects have been observed in near-IR spectroscopy by LP. We have adopted their values whenever appropriate (i.e. when the object was detected in their survey but not in ours or when they set a more stringent upper limit). Four objects have been detected in both surveys and can be used to assess the robustness of our luminosity calibration (this is in particular necessary when considering that some of our observations were acquired while some cirrus clouds were present; see Table 1). In all cases, both values are found to agree within the uncertainties. The internal consistency of our data can also be examined by comparing repeated observations of SNR28 taken under different photometric conditions, as detailed in Table 1. Once again, these quantities are found to be statistically undistinguishable. These checks confirm the reliability of our derived luminosities.  

The contour maps of the detected SNRs are shown in Fig. 2. Although the limited spatial resolution of our images prevents us to examine this aspect in detail, a comparison with Hubble Space Telescope ({\em HST}) or ground-based optical images (Blair \& Davidsen 1993; Gordon \etal 1998) reveals roughly similar [Fe II]- and optical-line morphologies. We note significant differences between the diameters determined from our \fe images and from optical observations (Gordon \etal 1998). This might result from difficulties in subjectively assigning diameters for objects with complex morphologies (see Blair \& Davidsen 1993). 

\begin{figure*}
\epsfxsize=14cm
\epsfysize=14cm
\epsfbox{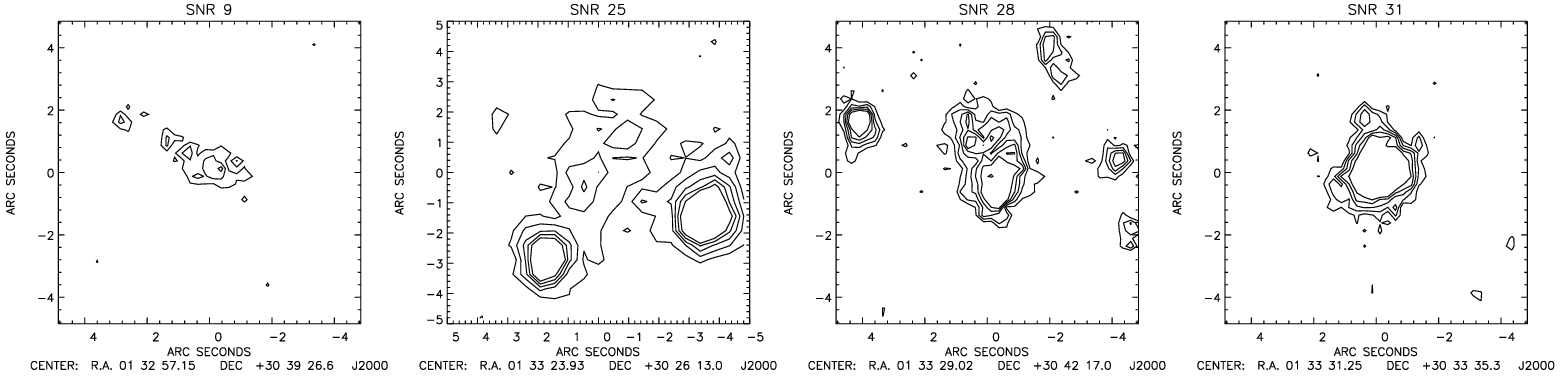}
\end{figure*}

\begin{figure*}
\epsfxsize=14cm
\epsfysize=14cm
\epsfbox{fig2b.ps}
\vspace*{-18.7cm}
\caption{Contour maps of the detected SNRs. The identification of the SNR (taken from Gordon \etal 1998) is indicated on top of each panel. The contours are drawn from 3$\sigma$ to 11$\sigma$, by steps of 2$\sigma$. At the assumed distance to M33, 1 arcsec corresponds to 4.1 pc. The field of view is 10 arcsec $\times$ 10 arcsec.
North is up and east is to the left.}
\end{figure*}

\section{Correlations between the \fe emission and other SNR properties}
In order to pin down the physical processes that are responsible for the \fe emission, we sought for correlations between the \fe luminosities and various quantities intrinsic to the SNRs (e.g. optical-line properties). These data were gathered from the literature and are quoted in Table \nolinebreak 2.\linebreak

A number of statistical methods have been developed to investigate relationships between astronomical data sets with censored data (e.g. with upper limits). We made use here of the generalized Kendall's $\tau$ correlation technique (Isobe, Feigelson \& Nelson 1986). Table 3 summarizes the results of this statistical analysis for all quantities under consideration. When a significant correlation is present (we consider this to be the case when the false alarm probability is less than 1 per cent), we give the results of a linear regression fit to the data for the {\em detected} objects (such regression techniques for censored data yield unreliable results in the presence of heavy censoring). We also show in Fig. 3 the dependence of \lfs on the quantity in question. Survival analysis techniques yield more reliable results when the censored points are randomly distributed, a condition that is generally not fulfilled here (see Fig.3). In order to assess the robustness of our results, we performed the same analysis when only considering firm detections. This procedure broadly confirms the correlations previously found (albeit with a lower level of significance), and suggests that the results inferred by taking into account all the available information (i.e. detections {\em and} upper limits) can be regarded as robust. We discuss below the implications of this statistical analysis on our understanding of \fe emission in SNRs.

\setcounter{table}{2}
\begin{table*}
\begin{minipage}{177mm}
\centering
\caption{Results of the statistical analysis. $N$: Number of points included in the calculations. ${\cal P} (X, Y)$: Probability for a chance correlation between the variables $X$ and $Y$ (data from Table 2). Note that the objects at large galactocentric distances, $GCD$, show a tendency for {\em lower} \lfs and $n_e$ values.} 
\hspace*{-1.0cm}
\begin{tabular}{ccrcrcl}
\hline
$X$  &  $Y$                                 & $N$     & ${\cal P} (X, Y)$ & $N$                      & ${\cal P} (X, Y)$ & \multicolumn{1}{c}{$\log X$ = ($a$ $\pm$ $\sigma_a$) $\log Y$ + ($b$ $\pm$ $\sigma_b$)} \\
     &  &                                    &  (per cent)       &                   & (per cent) & \\
     &                                      &         &                   & \multicolumn{2}{c}{(detections only)} & \multicolumn{1}{c}{(detections only)}\\
(1) & (2) & (3) & (4) & (5) & (6) & \multicolumn{1}{c}{(7)}\\\hline
\lf  & Dynamical age$^a$                    & 42      &  82             & 10                       & 100             & \\
 & $L$([O I] $\lambda$6300)   & 34      & 0.1             & 10                       & 1.6             & $\log$($L_{{\rm [Fe \: II]}}$) = (1.345 $\pm$ 0.167) $\log$($L$ [O I] $\lambda$6300) + (-- 0.402 $\pm$ 0.343)\\
 & $L$(H$\alpha$)             & 34      & 0.2             & 10                       & 2.5             & $\log$($L_{{\rm [Fe \: II]}}$) = (1.174 $\pm$ 0.159) $\log$($L$ H$\alpha$) + (-- 0.771 $\pm$ 0.422)\\
 & $L$([N II] $\lambda$6584)  & 34      & 0.1             & 10                       & 0.6             & $\log$($L_{{\rm [Fe \: II]}}$) = (0.949 $\pm$ 0.133) $\log$($L$ [N II] $\lambda$6584) + (0.210 $\pm$ 0.300)\\
 & $L$([S II] $\lambda$6717)  & 34      & 0.1             & 10                       & 0.2             & $\log$($L_{{\rm [Fe \: II]}}$) = (1.244 $\pm$ 0.165) $\log$($L$ [S II] $\lambda$6717) + (-- 0.577 $\pm$ 0.388)\\
 & $v_{\rm bulk}$                       & 8       & 8.3             & 8                        & 8.3             & \\
 & $n_e$                            & 34      & 0.1             & 9                        & 3.7             & $\log$($L_{{\rm [Fe \: II]}}$) = (1.244 $\pm$ 0.188) $\log$($n_e$) + (-- 0.590 $\pm$ 0.454)\\
 & $S_6$                            & 35      & 1.5             & 8                        & 11              &\\
 & $S_{20}$                         & 35      & 2.2             & 8                        & 14              &\\
 & $A^b$                              & 20      & 0.2             & 9                        & 0.7       & $\log$($L_{{\rm [Fe \: II]}}$) = (1.380 $\pm$ 0.216) $\log$($A$) + (7.342 $\pm$ 0.768)\\
 & $L_{\rm X}$                            & 8       &  53             & 4                        & 50              & \\
 & $GCD$                            & 42      & 8.5             & 10                       & 18              &                  \\
$n_e$    & $A^b$                      & 20      & 0.1             & 12                       & 1.1             &  $\log$($n_e$) = 1.753 $\log$($A$) + 8.530\\                 
    & $GCD$                    & 34      & 18              & 20                       & 52              & \\\hline
\end{tabular}
\begin{flushleft}
$^a$The dynamical ages have been derived using equation (5) of Gordon \etal (1998).\\
$^b$``Metallicity index''; see Section 5.3 for definition.
\end{flushleft}
\end{minipage}
\end{table*}

\begin{figure*}
\epsfxsize=14cm
\epsfysize=14cm
\epsfbox{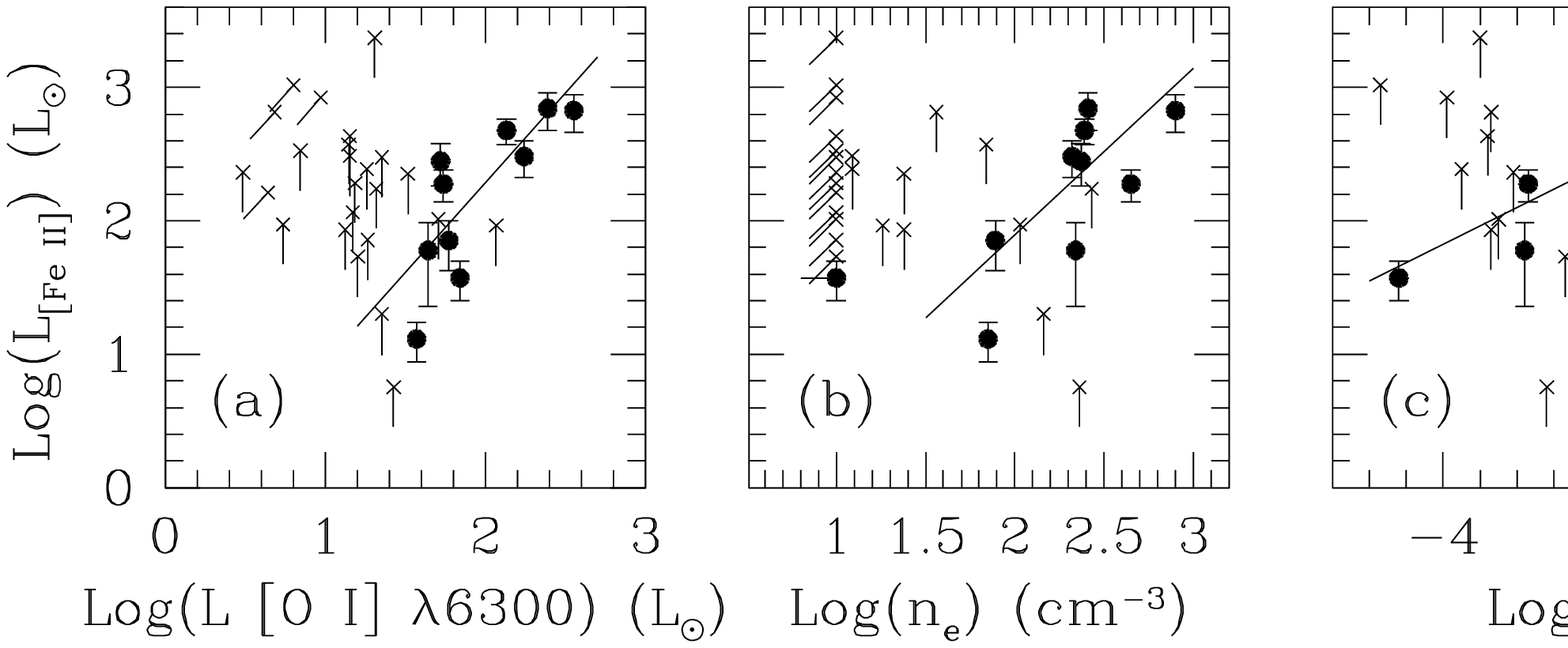}
\vspace*{-7.3cm}
\caption{Variation of \ls with ({\em a}) $L$([O I] $\lambda$6300), ({\em b}) $n_e$ and ({\em c}) the ``metallicity index'' $A$ (data from Table 2). These three quantities show a statistically significant correlation (${\cal P}$ $<$ 1 per cent) with \lfs (see Table 3). Filled circles are firm detections while the crosses denote 3$\sigma$ upper limits. The solid line shows the results of a linear regression fit derived from $\chi^2$ fitting of the data for the {\em detected} objects. Note that Fig.3{\em a} is meant to illustrate the correlation between the optical- and [Fe II]-line luminosities. Similar variations are observed for the other optical transitions.}
\end{figure*}

\subsection{Optical-line properties}
A significant correlation is found between all optical-line luminosities and \lfs (Table 3). A good correlation between the [Fe II] and H$\beta$ surface brightnesses has already been reported for a sample of LMC and Galactic SNRs by Oliva \etal (1989). Although the optical and \fe line-emitting regions are likely to differ in detail, this correlation suggests that the sites of \fe emission are bright optical filaments whose emission is excited by radiative shocks. Support for this interpretation comes from observations of RCW 103 and N49 which show the [Fe II]- and optical-line morphologies to be virtually identical (Burton \& Spyromilio 1993; Dickel \etal 1995). In both cases, the optical line-emitting regions are believed to trace the interaction of a radiative blast wave with a neighbouring molecular cloud (e.g. Banas \etal 1997; Oliva \etal 1999). 

\subsection{Density effects}
A correlation is also found between the electronic density of the postshock gas (as derived from the optical [S II] doublet) and \lf. Near-IR \fe features in SNRs are collisionally excited via electron impact in the dense recombination zone behind the shock front. The critical electron number density of [Fe II] $\lambda$1.644 $\mu$m is (Blietz \etal 1994): 
\begin{equation}
n_c \: (T_4) \approx 4.2 \times 10^4 \: T_4^{0.69} \hspace*{0.2cm} {\rm cm}^{-3}
\end{equation}
In this expression, $T_4$ is the electronic temperature in units of 10$^4$ K. For temperatures typical of \fe line-emitting regions in SNRs ($T_e$ $\approx$ 6 500 K; Oliva \etal 1989), collisional de-excitation becomes important for: $n_e$ $\approx$ 3.1 $\times$ 10$^4$ cm$^{-3}$. The postshock densities are of the order of 10$^3$ cm$^{-3}$ for the \fe brightest remnants (Fig.3$b$). However, the densities derived from the optical [S II] doublet have been generally found to be systematically lower (by about a factor of 5) than those determined from the near-IR \fe lines, suggesting that the \fe features are produced further downstream behind the shock (e.g. Oliva \etal 1989). The electronic densities prevailing in the [Fe II]-line emitting regions of the SNRs in our sample are thus less than  5 $\times$ 10$^3$ cm$^{-3}$, but are still much below the critical density for collisional de-excitation. In a two-level approximation, the luminosity of the \FES transition is given by (Blietz \etal 1994):
\begin{equation}
L_{{\rm [Fe \: II]}} = \pi^2 \: \frac{T_4^{-0.94} \: e^{-1.57/T_4} \: {\cal N_{{\rm Fe^{+}}}}}{1 + n_c(T_4)/n_e} \: \left(\frac{d}{\rm cm}\right)^2 \hspace*{0.2cm}  {\rm ergs \: s^{-1}}
\end{equation}
Here $\cal{N_{{\rm Fe^{+}}}}$ is the column density of Fe$^+$ in units of 10$^{16}$ cm$^{-2}$, $n_c$ is the critical density given by equation (2) and $d$ is the diameter of the SNR. In view of the similar size of the detected SNRs (see Fig.2), we will ignore in the following any dependence of \lfs on the latter quantity. While \lfs is only weakly dependent on the temperature (it varies within a factor of only 1.5 over 5 decades in density between 6 000 and 20 000 K), it can be expressed to a reasonable degree of accuracy as a linear function of $n_e$ for densities up to 10$^4$ cm$^{-3}$. The relationship found in our data between \lfs and $n_e$ (\lfs $\propto$ $n_e^{1.244 \pm 0.188}$) is close to what might be expected from theoritical considerations. We will return to this point in the following.

The postshock densities derived for the detected SNRs are 2 to 4 orders of magnitude higher than canonical values for the ISM ($n_0$ $\approx$ 0.1--10 cm$^{-3}$). We have previously argued that the \fe brightest SNRs are in the radiative expansion phase. Large compression factors (i.e. ratio of postshock to preshock density) can be achieved for strong radiative shocks, with values in the range 10--50 being typical (e.g. Vancura \etal 1992). While we cannot rule out the possibility that the high postshock densities inferred for the detected SNRs are due to very large compression factors, we rather favour the idea that they are evolving in an ambient medium of high density. The tendency for the objects with high radio fluxes to be strong \fe emitters supports this picture (Table 3). Since thermal X-ray emission from the shock-heated gas (which should dominate X-ray excitation in our sample) scales with ambient density (e.g. Magnier \etal 1997), X-ray data can also be used to assess the relevance of this interpretation. However, we find no evidence in our data for a relationship between the X-ray and the \fe luminosities. A better knowledge of the X-ray properties of SNRs in M33 might shed some new light on this issue. 

\subsection{Metallicity effects}
Abundances derived from modelling the optical-line properties of SNRs reflect to a large extent the chemical composition of the swept-up interstellar material (e.g. Russell \& Dopita 1990), and can be characterized by the 
``metallicity index'', $A$, defined as: $\log(A)$ = 0.8 + 1/3[$\log(N/H)$ + $\log(O/H)$ + $\log(S/H)$] (Dopita \etal 1984). We find a correlation between \lfs and $A$ which might indicate that \fe emission is enhanced in regions characterized by high ISM chemical abundances. The column density of Fe$^+$, ${\cal N}_{{\rm Fe^{+}}}$, in equation (3) can be rewritten as:
\begin{equation}
{\cal N}_{{\rm Fe^{+}}} = f_1 \: \delta \: X^{\odot}_{{\rm Fe}} \: {\cal N}_{\rm H} \hspace*{0.2cm}  {\rm cm}^{-2}
\end{equation}
where $f_1$ is the ionization fraction of Fe$^+$, $\delta$ is the gas phase iron abundance, $X^{\odot}_{{\rm Fe}}$ is the solar abundance of iron relative to hydrogen and ${\cal N}_{\rm H}$ is the column density of neutral and singly ionized hydrogen. It can be seen that the correlation found in our data between \lfs and $A$ is consistent with theoritical predictions. 

One major difficulty with this interpretation lies, however, in the existence of a positive correlation between $n_e$ and $A$ (see Table 3 and Fig.4). While there is no evidence in our data for a negative correlation between $n_e$ and galactocentric distance, this might suggest that both the cosmic abundances (see Smith \etal 1993) and the density of the ISM increase when progressing towards the inner parts of M33. In view of this unexpected result, it is not clear whether the correlation between \lfs and $A$ is fundamental or is mainly a density effect. It is conceivable, however, that a spread in the metallicity of the ambient ISM among our sample of SNRs contributes to the large scatter observed in the \lf-$n_e$ relation. Variations in the depletion factor might also be relevant in this respect, as both theory and observations suggest that the fraction of iron-bearing grains returned to the gas phase by shocks may significantly vary from one SNR to another (Jones, Tielens \& Hollenbach 1996; Oliva \etal 1999). The positive correlation between \lfs and the iron cosmic metal abundance hinted at by equation (4) should be kept in mind when comparing the \fe properties of SNR populations in host galaxies with widely different metallicities.

\begin{figure}
\epsfbox{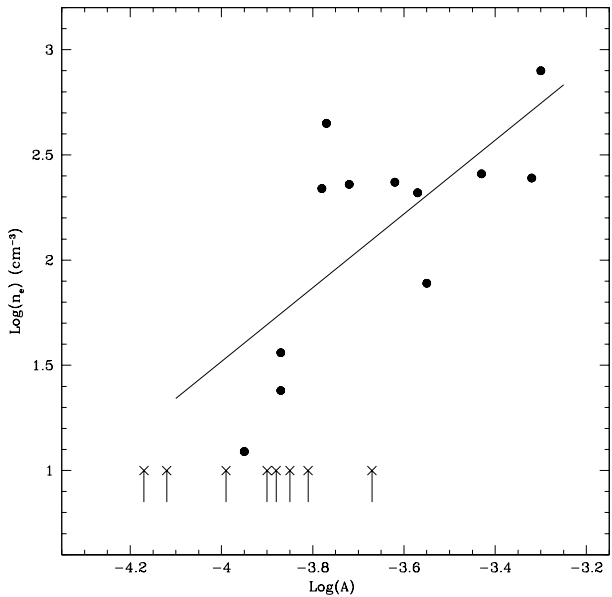}
\vspace*{-12cm}
\caption{Variation of $n_e$ with $A$. These two quantities show a statistically significant correlation (see Table 3). Filled circles are firm detections while the crosses denote 3$\sigma$ upper limits. The solid line shows the results of a linear regression fit derived from $\chi^2$ fitting of the points with a firm $n_e$ determination.}
\end{figure}

\section{Discussion}
\subsection{Near-IR \fe emission from SNRs}
In the light of the results presented above, we conclude that strong near-IR \fe emission in SNRs is subordinated to the existence of a radiative blast wave propagating through an ambient material of high density (other physical processes than shocks might account for \fe emission in very young SNRs; Graham \etal 1990). This picture is supported by the shock models of Hollenbach \& McKee (1989). The \fe brightest SNRs might be at the onset of the radiative phase, where line luminosities are believed to reach their maximum (Falle 1981; Cioffi \& McKee 1988). We note that the relatively small size of our detected SNRs (Fig.2) is not in conflict with the fact that they have reached such an advanced evolutionary stage; their confined nature may cause them to evolve rapidly. The low detection rate of our survey may result from: (i) the finite duration of strong \fe \linebreak emission in SNRs (see Section 6.4) and (ii) the fact that optically-selected samples of SNRs are biased in favour of objects evolving in a teneous medium, possibly the warm component of the ISM (e.g. Pannuti \etal 2000). 

The following empirical relation has been found in our data between \lfs and $n_e$ (see Table 3): \lfs (L$_{\odot}$) $\approx$ 0.257 $n_e^{1.244 \pm 0.188}$ (cm$^{-3}$). Since this exponent of 1.244 $\pm$ 0.188 is very close to the value of unity expected on theoritical grounds (see Section 5.2), we shall consider in the following that: 
\begin{equation}
L_{{\rm [Fe \: II]}} = (1.1 \pm 0.3) \: \left(\frac{n_e}{{\rm cm}^{-3}}\right) \hspace*{0.2cm}  {\rm L_{\odot}}
\end{equation}
The factor 1.1 is simply the weighted average of \lf/$n_e$. The equation above can be used to estimate the \FES line luminosity of {\em radiative} SNRs from the sole knowledge of the electronic density in the postshock region (since this relation refers to the density in the [S II]-emitting region, care should be taken when using other density diagnostics). We caution, however, that this relation has been derived on the basis of few detections that only sample a limited density regime. A better knowledge of the \fe properties of SNRs in host galaxies with widely different ISM properties is needed to fully assess the robustness of this relationship, especially with respect to metallicity effects. 

\subsection{Comparison with Galactic and LMC SNRs}
Is this picture consistent with \fe observations of SNRs in our Galaxy and in the LMC? These observations are summarized in Table 4, along with the typical postshock densities associated with the SNRs considered. Great care has been taken to ensure that the density estimates (as determined from the [S II] doublet) relate to the same regions observed in [Fe II]. One major difficulty in comparing the \fe luminosities of SNRs in M33 with their counterparts in our Galaxy or in the LMC lies in the extended nature of the latter class of objects. Most \fe luminosities quoted in Table 4 are derived from aperture photometry and, in most cases, only sample a very small fraction of the SNR line-emitting regions. We performed an aperture correction to estimate the total \fe luminosities by considering the ratio of the total optical size of the remnant by the area sustained by the aperture used.\footnote{We note that these corrections are likely to be grossly overestimated since the \fe measurements are generally performed on the brightest optical knots where we expect the highest level of \fe emission (see Section 5.1). For instance, we derive from the observations of Rho \etal (2001) a total dereddened \FES luminosity for IC 443 of about 55 L$_{\odot}$. This value is much smaller than our estimate of 1 200 L$_{\odot}$ derived from the measurements of Graham \etal (1987) through a 35 arcsec $\times$ 35 arcsec aperture.} We show in Fig. 5 the resulting \fe luminosities of SNRs in M33, our Galaxy and the LMC, as a function of $n_e$. A reasonable agreement between the expected and observed \fe luminosities of IC 443, the Cygnus Loop, N49, RCW 103 and N63A is found considering the crudeness of the corrections applied. Interestingly, the 3 SNRs (Crab, Kepler and N103B) that fail by more than 2 orders of magnitude to follow the relationship between \lfs and $n_e$ are the objects in our sample of Galactic or LMC SNRs most unlikely to undergo radiative expansion. The Crab nebula and Kepler's remnant are both very young and therefore are still in the free (or Sedov-Taylor) expansion phase. The same is true for N103B which is likely to be still undergoing adiabatic expansion (Dickel \& Milne 1995). The weakness of the \fe emission in N103B and in Kepler's remnant illustrates the important point that the existence of radiative shocks is a necessary condition for strong \fe emission, but {\em not} a high-density ambient ISM. As can be seen in Fig.3{\em b}, the \fe luminosities of SNRs with similar densities may differ by as much as 2 orders of magnitude. This is best explained by evolutionary effects, with the low \fe emitters being dominated by adiabatic shocks.   

\begin{table*}
\begin{minipage}{177mm}
\centering
\caption{\lfs and postshock electronic densities for LMC and Galactic SNRs.} 
\begin{tabular}{lcccccccc}
\hline
Object     & $D$   & Optical size            &  \lf$^a$                 & Aperture size             & Ref. & $n_e$([S II])                  & Aperture$^b$          & Ref.$^c$\\
                               & (kpc) & (arcsec $\times$  arcsec)    & (L$_{\odot}$)            & (arcsec $\times$ arcsec) &      & (10$^{3}$ cm$^{-3}$) & ID              &  \\
(1)        & (2)   & (3)                     & (4)                      & (5)                  & (6)  & (7)                            & (8)                 & (9) \\\hline
N63A                           & 52    & 25 $\times$ 25          & 240                      & 20 $\times$ 30       & 1    & 1.3                            & --                 & 2 \\
N49                            & 52    & 65 $\times$ 65          & 720                      & 20 $\times$ 35       & 1    & 0.6                            & {\em d} and {\em e} & 3 \\
N103B                          & 52    & 25 $\times$ 25          & 26                       & 20 $\times$ 15       & 1    & 6.0                            & --                 & 2 \\
RCW 103                        & 6.0   & 340 $\times$ 560        & 170$^d$                  & 75 $\times$ 30       & 1    & 1.0                            & 2                   & 4 \\
MHS 15--52                     & 4.2   & 2 100 $\times$ 2 100$^e$  & 1.0                      & 5 $\times$  5        & 5    & 1.7                            & --                 & 5 \\
Kepler                         & 4.0   & 21 $\times$ 64          & 0.7                      & 15 $\times$ 15       & 1    & 6.0                            & 1 and 2$^f$         & 4 \\
Crab                           & 2.0   & 290 $\times$ 420        & 0.3$^g$                  & 75 $\times$ 75       & 1    & 0.6                            & 3--8 and 10          & 6\\
IC 443                         & 1.5   & 2 700 $\times$ 2 700$^e$  & 0.2$^h$                      & 35 $\times$ 35       & 7    & 0.1                            & 1--3                 & 8 \\
Cygnus Loop                    & 0.44  & 13 800 $\times$ 9 600$^e$ & 9 $\times$ 10$^{-4}$     & 20 $\times$ 20       & 9    & 0.2                            & 1                   & 10\\
 \hline
\end{tabular}
\begin{flushleft}
$^a$[Fe II] $\lambda$1.644 $\mu$m luminosities corrected for extinction. \\
$^b$Apertures where the [S II] ratios have been determined. The mean of the [S II] ratios has been considered when several aperture IDs are given.\\ 
$^c$[S II] ratio from these references; the electronic densities have been derived using the diagnostic diagram (see their fig.6) of Blair \& Kirshner (1985).\\
$^d$See also Oliva \etal (1999) for measurements through smaller apertures.\\
$^e$From Green (2000).\\
$^f$Corresponds to knot \#27 of D'Odorico \etal (1986).\\ 
$^g$See also Graham \etal (1990), Hudgins, Herter \& Joyce (1990) and Rudy, Rossano \& Puetter (1994) for measurements of near-IR \fe line fluxes through smaller apertures.\\
$^h$Because of the existence of noticeable density gradients across this SNR (and the related difficulty in assigning a typical density), the total value of Rho \etal (2001) is not considered here.\\
References -- (1): Oliva \etal (1989); (2) Danziger \& Leibowitz (1985); (3) Vancura \etal (1992); (4) Leibowitz \& Danziger (1983); (5) Seward \etal (1983); (6) Fesen \& Kirshner (1982); (7) Graham \etal (1987); (8) Fesen \& Kirshner (1980); (9) Graham \etal (1991); (10) Miller (1974).
\end{flushleft}
\end{minipage}
\end{table*}

\begin{figure}
\epsfbox{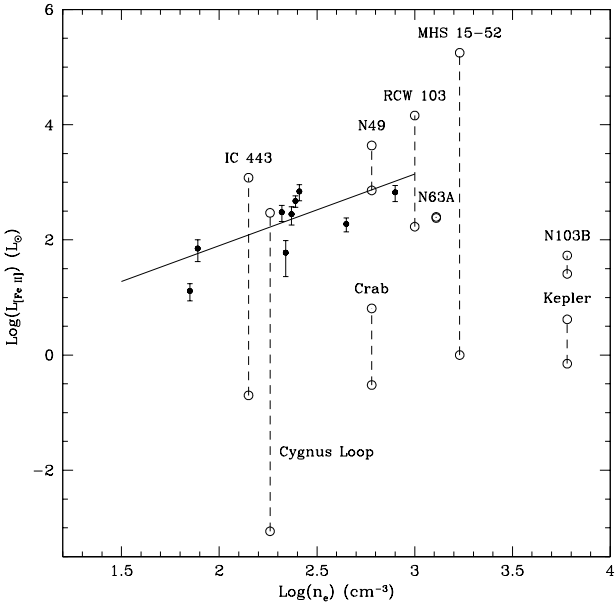}
\vspace*{-12cm}
\caption{Variation of \lfs with $n_e$ for the detected SNRs in M33 ({\em filled dots}) and for a sample of SNRs in our Galaxy and in the LMC ({\em open dots}). The dashed lines illustrate the aperture corrections applied to the \fe luminosities of the latter class of objects (see text). The solid line shows the results of a linear regression fit derived from $\chi^2$ fitting of the SNRs in M33 with a firm $n_e$ determination.}
\end{figure}   

\subsection{Implications for the \fe emission of SNRs in starburst galaxies}
The fact that the level of \fe emission scales with density may help to understand the differences that exist between the near-IR \fe properties of SNR populations in galaxies with moderate star formation activity (such as M33) and starburst galaxies. While the present study demonstrates that the brightest remnants in M33 have \fe luminosities of the order of 700 L$_{\odot}$, SNRs in the starburst galaxy M82 have luminosities up to 1.6 $\times$ 10$^5$ L$_{\odot}$ (Greenhouse \etal 1997). Much higher values are also observed in NGC 253 (Forbes \etal 1993). We compare below the \fe luminosities expected within the framework of the results presented above with the observed values. 

The electronic density in the [Fe II]-emitting regions of NGC 253 has been directly measured from density-sensitive near-IR \fe line ratios and varies from 5 $\times$ 10$^3$ cm$^{-3}$ (Engelbracht \etal 1998) to 1 $\times$ 10$^4$ cm$^{-3}$ (Simpson \etal 1996). The [Fe II] emission in NGC 253 is believed to mainly arise from SNRs in the disk of the galaxy, and not from superwind- or AGN-related phenomena (Forbes \etal 1993). It is therefore reasonable to assume that these densities are typical of the postshock regions of SNRs in the central regions of this galaxy. Keeping in mind that our empirical relation (equation 5) relates to the density derived from the [S II] doublet and that the density in the [Fe II]-emitting region is about 5 times higher (e.g. Oliva \etal 1989), we estimate for these SNRs: \lfs $\approx$ 1.1--2.2 $\times$ 10$^3$ L$_{\odot}$. This is in good agreement with the observed value: \lfs $\approx$ 2.6 $\times$ 10$^3$ L$_{\odot}$ (Forbes \etal 1993).

The ISM electronic densities prevailing in the central regions of M82 show a large spread: $n_0$ = 50--500 cm$^{-3}$ (F\"orster-Schreiber \etal 2001, and references therein). We shall consider in the following: $n_0$ =  200 cm$^{-3}$. This translates into: $n_e$ $\approx$ 6 $\times$ 10$^3$ cm$^{-3}$, if one considers a canonical value for the compression factor of the SNR blast wave of 30. This estimate is consistent with the upper limit of 3.2 $\times$ 10$^4$ cm$^{-3}$ derived by Lester \etal (1990) from \fe lines. According to equation (5), \FES luminosities approaching 10$^4$ L$_{\odot}$ can easily be reached for these SNRs. However, it appears that density effects alone have difficulties in explaining the high luminosities of the \fe brightest sources in M82 (up to 1.6 $\times$ 10$^5$ L$_{\odot}$; Greenhouse \etal 1997). Two effects might account for this discrepancy. First, the SNR might expand in the interclump medium of molecular clouds, with the result that we severely underestimate $n_e$ (Chevalier \& Fransson 2001). Second, in regard of the high space density of SNRs in the compact star-forming regions of this galaxy (Huang \etal 1994), it is far from clear that the strong \fe sources mapped by Greenhouse \etal (1997) can be identified with individual SNRs. The radius at which a SNR enters the radiative phase scales with the initial blast energy, $E_{51}$, and upstream hydrogen number density, $n_{\rm H}$, as: $R_{{\rm rad}}$ $\propto$ $E_{51}^{5/17}$$n_{\rm H}^{-7/17}$ (Blondin \etal 1998). Regardless of the details of the original explosion, one therefore expects the bright \fe SNRs in starburst galaxies to be less spatially extended (by a factor 2--5) than the detected SNRs in M33 (with $\overline{d}$ $\approx$ 21 pc). It is thus plausible that the exceptionally strong \fe sources in M82 (with sizes up to 40 pc) are made up of several individual SNRs with diameters in the range 4--10 pc.

We conclude that the dichotomy between the \fe properties of SNRs in quiescent and in starburst galaxies may be interpreted as a result of the different ISM densities prevailing in both type of galaxies. Although clear metallicity effects have not been uncovered from our data, we also note that a substantial enrichment in metals is expected in the hostile environments of starburst regions (because of high chemical abundances and/or processing via grain destruction of the large reservoir of dust grains) and might also potentially contribute to strong \fe emission. 

\subsection{The [Fe II]-emitting lifetime of SNRs}
To estimate the typical [Fe II]-emitting lifetime of a SNR, we consider the integrated \FES luminosity corrected for extinction in the central 2.4 arcsec $\times$ 12 arcsec region of NGC 253: $L_{{\rm [Fe \: II]}}$ $\approx$ 1.3 $\times$ 10$^5$ L$_{\odot}$ (Engelbracht \etal 1998). A distance of 2.5 Mpc has been adopted (De Vaucouleurs 1978). We have also assumed a homogeneous mixture of gas and dust with an extinction towards the [Fe II]-emitting regions of $A_V$ = 8.4 mag (Engelbracht \etal 1998) and the interstellar extinction law of Rieke \& Lebofsky (1985). The {\em global} radio supernova rate of NGC 253 is uncertain, but estimates range from 0.1 to 0.2 yr$^{-1}$ (Ulvestad 2000, and references therein). A value $\eta$ $\approx$ 0.15 yr$^{-1}$ is adopted in the following. To translate this into a normalized value for the central 2.4 arcsec $\times$ 12 arcsec region, we consider the ratio of the area sustained by this aperture by the total radio extent of the galaxy, which gives: 9 $\times$ 10$^{-3}$ yr$^{-1}$. The electron density in the [Fe II]-emitting regions lies in the range: 5--10 $\times$ 10$^3$ cm$^{-3}$ (Simpson \etal 1996; Engelbracht \etal 1998). Using equation (5)  by converting to densities in the [S II]-emitting regions, and substituting the result in equation (1) yields: $t_{{\rm [Fe \: II}]}$ $\approx$ 0.7--1.4 $\times$ 10$^4$ yr.

Similar calculations can be applied to M82. The integrated \FES luminosity corrected for extinction in the central 16 arcsec $\times$ 10 arcsec region is: $L_{{\rm [Fe \: II]}}$ $\approx$ 2.7 $\times$ 10$^6$ L$_{\odot}$ (F\"orster-Schreiber \etal 2001). A distance of 3.3 Mpc (Freedman \& Madore 1988) and the similar extinction model as above with $A_V$ = 36 mag has been adopted (F\"orster-Schreiber \etal 2001). By considering this relatively small area in M82, we minimize the contribution of large-scale, superwind-induced emission (Greenhouse \etal 1997). Radio observations yield a global supernova rate: $\eta$ = 0.11 $\pm$ 0.05 yr$^{-1}$, which translates into a normalized value for the central 16 arcsec $\times$ 10 arcsec region of about 0.016 yr$^{-1}$ (Huang \etal 1994). Additionally, we assume the typical \fe luminosity for SNRs in M82 determined previously: \lfs $\approx$ 10$^4$ L$_{\odot}$. Substituting in equation (1) yields: $t_{{\rm [Fe \: II}]}$ $\approx$ 1.7 $\times$ 10$^4$ yr.

In regard of the large uncertainties involved, the values of $t_{{\rm [Fe \: II]}}$ for SNRs in NGC 253 and M82 appear to be in good agreement. The [Fe II]-emitting lifetime derived for NGC 253 is more robust, as the density of the [Fe II]-emitting gas has been directly measured in this galaxy. A value of 1 $\times$ 10$^4$ yr will thus be adopted in the following. This estimate is broadly consistent with the duration of strong optical emission in SNRs, as determined by evolutionary models (e.g. Franco \etal 1994).

\subsection{The supernova rate of starburst galaxies}
By using equation (5) and the [Fe II]-emitting lifetime derived above, the supernova rate given by equation (1) can be written as a function of the {\em postshock} electron density as:
\begin{equation}
\eta  \approx \frac{{\cal L}_{{\rm [Fe \: II]}}/{\rm L_{\odot}}}{1.1 \times 10^4}\left(\frac{n_e}{{\rm cm}^{-3}}\right)^{-1} \hspace*{0.2cm} {\rm yr^{-1}}
\end{equation}
In the absence of density diagnostics for the postshock regions, this expression can be conveniently expressed as a function of the (more easily measurable) ISM number density, $n_0$, by an appropriate choice of the typical compression factor of the SNR blast wave.

In regard of the various assumptions made here and the fact that metallicity issues still remain to be fully addressed, the supernova rate we derive is still uncertain. In spite of these limitations, we note that our expression is consistent with previous estimates (van der Werf \etal 1993; Vanzi \& Rieke 1997) if canonical ISM densities for starburst galaxies are considered ($n_0$ $\approx$ 100 cm$^{-3}$). The relation proposed by van der Werf \etal (1993) has been found to give a good agreement in a sample of starburst galaxies between supernova rates derived from near-IR \fe lines and from their star-forming properties (Calzetti 1997). This provides some support to the validity of equation (6) to provide robust estimates of the supernova rates in galaxies with widely different ISM properties and for which the contribution of superwinds- and AGN-related phenomena to the total \fe output can be neglected. Because only radiative SNRs significantly contribute to ${\cal L}_{{\rm [Fe \: II]}}$, however, these values are only lower limits.  

\section{Conclusions}
The main conclusions of this study are the following:
\begin{itemize}
\item
We have carried out the first near-IR \fe line-imaging survey of extragalactic SNRs, presenting evidence for large intrinsic differences in their \fe properties.
\item
A significant correlation is found between the optical- and [Fe II]-line luminosities, suggesting that the detected SNRs are dominated by radiative shocks.
\item
We suggest that the SNRs with strong \fe emission are evolving in a dense ISM. Density effects are most likely the cause of the very strong level of \fe emission from SNRs in starburst galaxies.
\item
The typical [Fe II]-emitting lifetime of a SNR in starburst galaxies is found to be of the order of 10$^4$ yr.
\item
We provide a new empirical expression for the supernova rate of starburst galaxies, as derived from their integrated near-IR \fe luminosity. 
\end{itemize}  

\section*{Acknowledgments}
We wish to thank an anonymous referee for useful comments. We acknowledge the TACs of the CFHT for their generous time allocation. The Canada-France-Hawaii Telescope is operated by the National Research Council of Canada, the Centre Nationale de la Recherche Scientifique of France, and the University of Hawaii. The authors also wish to thank the Natural Sciences and Engineering Research Council (NSERC) of Canada and the Fonds pour la Formation de Chercheurs et l'Aide \`a la Recherche (FCAR) of Qu\'ebec for financial support. T. M. would like to thank the organizers of the conference ``The interstellar medium in M31 and M33'' (where the first results of this project were presented) for this enjoyable meeting, as well as the Wilhelm und Else Heraeus-Stiftung fundation for financial support.

\bsp

\label{lastpage}

\end{document}